# Infrared Microscopy of Biochemistry and Metabolism in Single Living Eukaryotic Cells

*Luca Quaroni*

*Jagiellonian University, Faculty of Chemistry, 30-387, Cracow, Poland*

*E-mail: luca.quaroni@uj.edu.pl*

**Abstract**

The turn of the millennium has seen a growing interest in the study of live cells by infrared (IR) spectroscopy, driven by the versatility, wealth of molecular information, and potential for high-throughput screening of the technique. Measurements on individual cells, either isolated or within a multi-cellular structure, provide information that is not available from ensemble samples. The present review discusses the use of infrared (IR) microscopy to analyse live single cells from a biochemical perspective, seeking information on real-time processes. The emphasis is on the use of the technique to quantify metabolic turnover, with the aim of providing a complementary method for metabolomics, and for toxicological and pharmacological studies. The work highlights the methodological advances and proof-of-concept experiments that took place over the past few years in this direction. It discusses current advantages and limitations of the technique, including the possibility of detecting specific biomolecules and their reactivity, and it concludes with a brief outline of future perspectives.

## *Introduction: Mid-infrared absorption spectroscopy in cellular biology*

The claim of IR spectroscopy as a tool for biochemical investigations in living cells relies on a few characteristics. All molecules, except for homonuclear diatomics, absorb light in the IR spectral range and all of them can be simultaneously probed. *A priori* selection of a molecular target or labelling with a probe are unnecessary, which qualifies it as an untargeted and non-invasive technique. Measurements are intrinsically non-destructive and can be performed on delicate samples with minimal perturbation. Furthermore, the technique can be applied in a microscopy optical configuration, to probe samples smaller than 100 μm. IR spectroscopy is sometimes described as a "discovery technique", because of its capability to disclose unexpected molecules and processes. These properties have proven useful in the study of dynamic processes in living or functional cells and tissue, and emphasis is shifting to measurements of single cells. The latter are driven by multiple opportunities, such as isolating cellular properties from those of the surrounding medium or matrix, studying cell-to-cell variability, identifying rare phenotypes in extended populations, measuring structure and dynamics in specific subcellular locations, and quantifying concentration gradients in space and time, just to mention a few. The subject has been reviewed a few years ago (Quaroni 2011a), and the present work extends the analysis to the current state-of-the-art.

IR spectroscopy relies on two empirical rules for the rapid characterization of sample composition. One is the recognition that the presence of certain functional groups in a molecule is associated with the absorption of light at specific wavelengths, and the observation of the corresponding spectral bands can provide qualitative information on molecular structure. The other is that the pattern of absorption bands is a characteristic *fingerprint* of a material, allowing the identification of unknowns. (Shurvell 2002) These rules have been the backbone of many applications in the study of soft matter, including biological samples. However, beyond these simple analytics, infrared spectra also provide a wealth of detailed molecular information. Absorption bands arise from energy transitions that excite molecular vibrations (vibrational modes). Band intensity and position are related to molecular geometry and to the strength of interatomic bonds, both of which can be extracted from spectral analysis. In reverse, the spectra of simpler systems can be calculated from known molecular properties, providing a way to confirm molecular identity and geometry. (Wilson 1980) While vibrational transitions occur over the whole IR spectral range, much attention is focused on the mid-infrared (mid-IR or MIR) spectral region (approx. 25 – 2.5 µm on the wavelength scale or 400 - 4000 cm$^{-1}$ on the energy-consistent wavenumber scale), where most molecular species, as well as many crystalline materials, have their fundamental vibrational transitions and absorb light strongly. The mid-IR spectral range will be the focus of this review and, unless otherwise specified, the terms "IR spectroscopy" and "IR microscopy" will refer to such range.

The usefulness of mid-IR spectroscopy in the analysis of biomedical samples was recognized with the introduction of the first commercial IR spectrometers. A major incentive has been the promise of diagnostic applications, which rely on the possibility to discriminate between healthy and diseased tissue by relying on the overall spectral trace as a marker for cellular identity and state. This topic falls beyond the scope of the present review and interested readers are directed to works focused on the specific subject (e.g. Diem 1999, Naumann 2001, Lasch 2002, Ellis 2006).

From a biochemical perspective, the appeal of mid-IR spectroscopy lies in the analysis of structure-function relationships at the molecular level. Spectral analysis provides molecular information, together with a quantitative assessment of concentrations. When the two are paired in time resolved measurements, kinetic and structural information can be simultaneously extracted, an approach which has been instrumental to many studies of biomolecules in bulk solution and has been extended to living cells. (Quaroni 2011a)

Infrared spectroscopy of living cells comes with major hurdles to overcome, foremost among them the compositional complexity of the samples. Thousands of different absorbing species are present, with concentration values that span a range from 10's M (water) to single molecules (DNA). The sum of these components provides an IR absorption spectrum that is qualitatively similar for most cells and tissues, such as the one of a hypothetical biological sample shown in Figure 1. Following convention, light absorption is expressed in absorbance units (Abs/au) and light frequency is given in wavenumbers ($\tilde{\omega}$ /cm$^{-1}$). Panel A shows the mid-IR region above 1000 cm$^{-1}$, frequently used for live cell spectroscopy. Panel B highlights part of the spectral region used for *fingerprinting*. Approximate locations of absorbance bands from some of the most abundant functional groups are also indicated. Characteristic strong bands, common to all cellular spectra, are typically observed in the 1700-1500 cm$^{-1}$ region, where the dominant contribution is from the amide groups of proteins, followed by absorption in the 2800-2950 cm$^{-1}$ region, from alkyl groups, and weaker contributions in the fingerprint region. Tabulated positions for the bands of the most relevant functional groups can be found in several reviews.

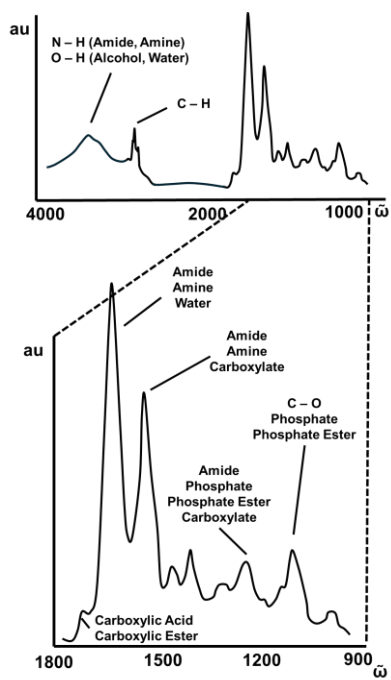

*Figure 1. Generic absorption spectrum of eukaryotic cells in the Mid-IR spectral region. A. Extended spectral region. B. Fingerprinting spectral region. The lines point to the approximate position of absorption bands characteristic of common functional groups.*

Despite the overall similarity of cellular spectra, important information is contained in small differences, which account for differences in chemical composition between cell and tissue types, and between individual cells and individual tissue samples. Extracting information from these spectra amounts to the core tasks of detection of the molecules of interest, their quantification, and the assessment of their distribution in time and space. The following sections will focus on different aspects of this challenge and provide some examples about how they can be addressed using infrared microscopy.

*IR spectroscopy of biological molecules and biochemical reactions*

IR spectroscopy has provided extensive information on the properties of biomolecules, ranging from small molecules to biopolymers, with much interest being directed to proteins, for which detailed structure-function relationships could be extracted. (Arrondo 1993; Siebert 1995). The IR spectrum of a protein in the region of amide bond absorption is dominated by contributions from the backbone polypeptide chain, which is sensitive to the secondary structure. Helices, sheets and turns absorb light at characteristic wavelengths, and careful analysis of amide absorption bandshapes allows estimates of their relative abundance. The analysis frequently relies on the band termed Amide I at ~1620-1700 $cm^{-1}$, mostly a stretching vibration of the carbonyl group, although the Amide II and Amide III bands can also be used. (Arrondo 1993) When performed with polarized light, these measurements provide the orientation of secondary structural elements in anisotropic samples, such as membrane embedded proteins. (Tamm 1997) The IR absorption of amino acid residues and cofactors has also received attention, because of their role in determining structure and reactivity of proteins. (Barth 2000) Since individual residues or cofactors, as well as substrates and products, provide minor contributions to the overall protein absorption spectrum, identifying their contribution is a core challenge of biochemical IR spectroscopy.

Over the past fifty years instrumentation for IR spectroscopy has been dominated by Fourier Transform Infrared (FTIR) technology, and more so in biochemical IR spectroscopy. Light from a broadband thermal source is analysed by an interferometer, typically a Michelson design, and the spectral information is extracted via Fourier Transform of the resulting interferogram. Spectral multiplexing allows rapid and sensitive measurements over a broad spectral range, and a major contribution of FTIR instruments has been the implementation of time-resolved measurements of dynamic events. Time-resolved measurements are now routinely used in the study of structure-function relationships in proteins, most notably to characterize the mechanism of action of enzymes and transporters, and the folding/unfolding of the polypeptide chain. (Fabian 2002, Barth 2003) Changes in the absorption spectra of substrates, intermediates, products, inhibitors, and side chains can be assigned to specific molecular events. The formation and breaking of covalent and hydrogen bonds, ligand substitution in the coordination sphere of metal cations, protonation and deprotonation of acidic and basic groups, conformational changes, oxidation and reduction of cofactors can all be tracked from mid-IR absorption spectra. By following spectral evolution at relevant wavelengths, structural and compositional changes can be expressed quantitatively in terms of turnover rates, and correlated to kinetic measurements, providing valuable information for our understanding of biochemical reactions. Rapid time-resolved measurements are normally reported using difference spectra, which show only the spectroscopic changes from the start of a process. In aqueous media the constant contribution from bulk water does not appear in the difference, which highlights only the spectroscopic variations of interest.

*Mid-infrared microscopy*

The selective measurement of single cells relies on IR microscopes. The instruments follow the familiar design of optical microscopes in terms of ergonomics and optics, and include parallel optical paths for visible and mid-IR light (Figure 2). The main difference with respect to optical microscopes is the exclusive use of reflective optics for the mid-IR light path, which avoids IR light absorption by common optical materials and reduces aberrations. IR light is analysed by an interferometer for FTIR spectroscopy before entering the microscope and is measured by a detector after the sample. The visible and infrared optical paths are paraxial and parfocal, aligned to ensure that visible light images can be used to simultaneously inspect the sample and position the IR beam. The ratio of the intensity crossing the sample and the intensity crossing a reference provides the absorption spectrum. (Sommer 2002)

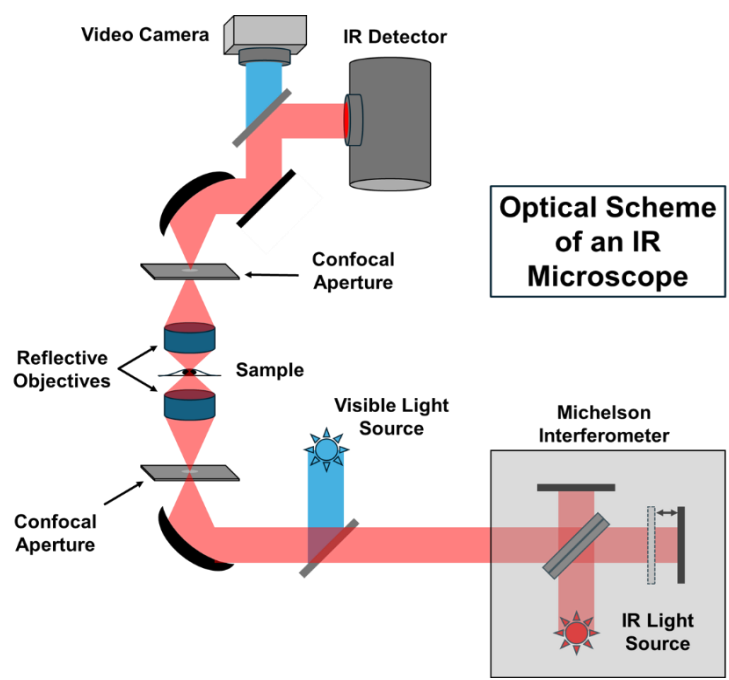

*Figure 2. Optical Scheme of an IR microscope. Optical scheme of an IR microscope aligned for transmission spectromicroscopy measurements. The light coming from the IR source is spectrally analysed by an interferometer. The modulated beam is focused on the sample using a reflective objective and collected by another objective on the opposite side. A set of confocal apertures define beam size and measured area. Light transmitted by the sample is collected and brought to a detector for intensity measurements. A dichroic mirror allows sample inspection by using visible light coaxial with the infrared beam.*

The simplest application, a spectromicroscopy (or microspectroscopy) measurement, uses the objectives of the microscope as beam condensers, to focus the IR beam into a small spot, which allows collecting absorption spectra from single microscopic locations, either in transmission or reflection. One or two confocal apertures limit the size of the IR beam and define the spatial resolution of the measurement. As in optical microscopy, the best available resolution, irrespective of aperture size, is set by optical diffraction, according to Rayleigh's criterion. Common IR microscope objectives have numerical aperture values of 0.4-0.6, and diffraction limited spatial resolution can be approximated by the wavelength of light, which is optimal for the spectroscopic investigation of single eukaryotic cells. The use of FTIR technology has provided a boost to the performance of IR microscopes because of its multiplexing capabilities, which allow simultaneous measurement of extended spectral regions with improved signal-to-noise ratio. The resulting detection limits are in the range of 10 - 100 pg for a spectromicroscopy measurement, suitable for measurement of a eukaryotic cell. The light source can be the standard thermal source of the FTIR spectrometer or an external source, such as light from a synchrotron. In the mid-IR spectral range, synchrotron radiation provides higher brightness when compared to conventional thermal sources, giving a gain in throughput and improved signal-to-noise ratio for diffraction limited measurements when coupled to commercial IR microscopes. The "brightness advantage" has drawn major interest for biomedical applications of IR microscopy (Moss 2010, and individual articles therein) and also promised faster measurements of single live cells with improved S/N ratio. (Quaroni 2011a, Mattson 2013)

In analogy with visible light microscopy, two dimensional images of the sample can be obtained either by raster scanning the focused beam over the sample (a practice often called mapping) (Wetzel 1999) or by expanding the beam and refocusing the image of the extended sample on a two-dimensional detector (a practice called imaging). (Lewis 1995) The dataset collected by mapping or imaging with an FTIR instrument is a spectral hypercube, with two axes representing the X and Y dimensions of the image and one representing the wavelength or frequency. A section of the cube at a given wavelength is an image of the sample at that wavelength, while the vector of data points at coordinates XY provides the absorption spectrum at that position. In an extension to three dimensions, tomographic imaging has been implemented by collecting tilt series of images while rotating the sample in the beam, followed by image reconstruction. (Martin 2013, Quaroni 2015)

Image contrast relies on intrinsic light absorption at a wavelength and does not require any staining. By matching this wavelength to the absorption peak of a species of interest, the resulting IR images or maps visualize its distribution throughout the sample, thus providing valuable compositional information. The capability of IR microscopy to generate contrast based on the intrinsic properties of a material, without the need for staining, and the rich molecular information contained in IR images justified the efforts devoted over the last three decades to the application of IR microscopy to biological samples, despite modest sensitivity and spatial resolution when compared to other widespread optical imaging techniques, such as fluorescence microscopy.

IR microscopy measurements of aqueous samples are most frequently carried out using either of two optical configurations, transmission or Attenuated Total Reflection (ATR). Transmission measurements commonly use sample holders based on a sandwich design, with two IR optical windows separated by a spacer and enclosing a thin layer of liquid or a humid atmosphere. The thickness of aqueous samples is minimized to reduce water absorption, and is typically in the range of a few micrometres, sufficient to enclose one or two layers of eukaryotic cells. Living cells are placed as a suspension between the two optical windows, or they are grown as adherent layers on one of them (Figure 3). The material of the optical windows must be insoluble in water and compatible with cellular biology, beside satisfying the spectroscopic requirements of the experiment. (Nasse 2009, Wehbe 2013, Mitri 2013) Glass and quartz can be used, but most of the fingerprint region is lost due to absorption by $SiO_2$. $CaF_2$ is most common, having the added advantage that its refractive index is a close match to that of water. Higher refractive index materials give higher numerical aperture in microscopy, but at the cost of forming optical interference fringes. (Chan 2018) Because of the thinness of the liquid phase, these holders are by definition microfluidic systems, and the more complex units require the use of specialized microfabrication techniques. In recent years, much of the effort in the field has been dedicated to the construction of more elaborate prototypes, to satisfy different experimental needs. (e.g. Nasse 2009, Marcsisin 2010, Tobin 2010, Vaccari 2012, Mitri 2013, Doherty 2019, Clède 2020)

In the ATR configuration the IR beam is internally reflected inside an optically transparent prism, while the sample is in contact with the reflecting surface. Optical interaction between sample and light is mediated by the evanescent field of the reflected wave established at the prism-water interface (Figure 3). (Fitzpatrick 2002) The evanescent field extends for about 0.2-10 µm into the aqueous phase, depending on the refractive index of the reflecting element, and is absorbed only by materials in contact with the surface, facilitating the measurement of adsorbates. An additional advantage is the ease of sample handling. Because the measurement probes only the volume within the evanescent field, there are no restrictions on the thickness of the sample, and aqueous samples can be conveniently contained in an open chamber. The configuration is particularly advantageous for adherent cell layers, the spectrum of which can be measured selectively from that of the medium, in contrast to suspended cells. ATR measurements can be extended to single cells and small tissue fragments by mounting ATR crystals, typically hemispheres or truncated cones, in the focal point of IR objectives (ATR objectives). (Fitzpatrick 2002) ATR crystals for this application are made of high refractive index materials that allow cellular growth or viability, typically Si, Ge, diamond and ZnS. (Chan 2016) Commercial setups for micro-ATR spectroscopy can be used, although the design mount may have to be modified to facilitate the handling of living samples and avoid pressure which may damage them. (Massaro 2008, Kuimova 2009)

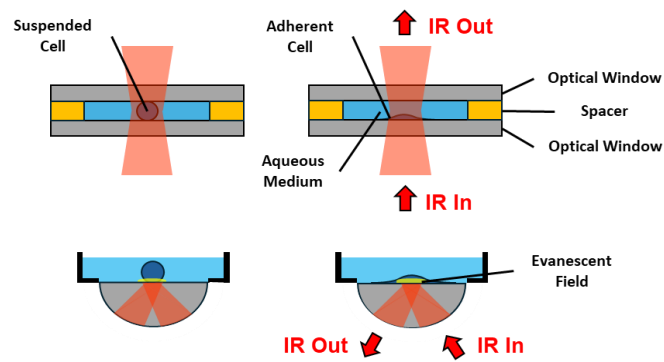

*Figure 3. Sample Holders for IR Microscopy of Aqueous Cells. Basic configurations of sample holders for IR microscopy of aqueous eukaryotic cells. Upper: Sandwich configuration for transmission measurements. Two IR transparent optical windows are separated by a micrometric spacer and contain the aqueous sample of cells. The IR beam probes the cell and the medium in its immediate surroundings, including the headspace above adherent cells. Lower: Configuration for micro-ATR measurements of cells. A spheroid or truncated cone is used as the ATR crystal. The evanescent field (in yellow), formed by internal reflection at the crystal–solution interface, extends through a thin interfacial layer of the sample and light is absorbed by cells within this layer. A substantial portion of the volume of an adherent cell is probed by the evanescent field, whereas only a small portion of a suspended cell is probed.*

*IR Microscopy of single living cells*

The dimensions of most prokaryotes and archaea are below the diffraction limit of mid-IR light, which severely constrains measurements on single cells. Spectra of single prokaryotic cells are dominated by the absorption of the surrounding medium or matrix, and the contribution from the cell itself is weak, severely limiting dynamic studies. Consequently, IR microscopy experiments on prokaryotes are normally performed on macroscopic cell ensembles. (Naumann 2001) In contrast, eukaryotic cells, ranging from a few micrometres to a hundred micrometres, prove to be an optimal complement to the performance of IR microscopes. Diffraction limited spatial resolution in the mid-IR is adequate for single cell and, in some cases, subcellular measurements. To date, most investigations on single cells involve dried or fixed samples and are aimed at the discrimination and classification of population subsets using the whole spectral trace as a marker of cellular composition. Analysis is concerned with reporting the collected contribution from functional groups and from some abundant classes of biomolecules and aims to identify and classify cells according to phenotype and cellular state (Diem 1999, Naumann 2001, Ellis 2006), and is often complemented by Raman microscopy measurements. (Salzer 2009)

The elevated water content constitutes a hurdle in transitioning from dried cell to live cell measurements and is the main constraint in sample preparation and handling for this otherwise versatile technique. Water absorption dominates the spectroscopic properties of the sample and limits sample thickness. Maximum thickness values are often taken to be ~1-2 μm for measurements close to 3400 $cm^{-1}$ and ~10-15 μm for measurements close to 1645 $cm^{-1}$, to avoid saturation of the corresponding water bands. (Quaroni 2011a) For accurate quantitative measurements, even thinner samples need to be used, to ensure the applicability of Lambert-Beer law. While bulk water bands can be digitally subtracted, this does not remove the high noise levels associated to strong absorption bands. It also does not remove the contribution from intracellular, interstitial and hydration water. The spectroscopic properties of these forms of confined water can differ from the bulk (e.g. Brubach 2001; Ebbinghaus 2006) and their contribution to Mid-IR spectra of cells and tissue still needs to be clarified. In addition, absorption from other components of the medium is also a potential interferent, particularly when a thick headspace is allowed. Media of minimal composition, such as isotonic NaCl, can be used to remove this contribution (Zhao 2010), provided that the effect on cellular physiology is tolerable. Failing this, techniques for spectral resolution must be resorted to, to extract the spectral contribution specific to the cell, as described in later sections.

On the positive side, a distinct advantage of measurements in an aqueous medium is the possibility to reduce spectral distortions caused by discontinuities in the refractive index of the sample. (Quaroni 2011a). Distortions are attributed to scattering, refraction and interference effects at the interface between materials of different refractive index, most notably at the air sample interface. They can be so severe as to preclude qualitative spectral analysis and spectral classification. At the very least, they can prevent quantitative measurements. (e.g. Diem 1999) Biological materials have infrared refractive index in the 1.3 – 1.6 range, with the higher values observed for biominerals and solid aggregates, (Aas 1996) while for aqueous solutions it is in the 1.32 - 1.34 range (Venyaminov 1997), and sample submersion greatly reduces the refractive index discontinuity between sample and environment.

Early measurements of single intact and living cells concentrated on sample management, i.e. in ensuring cellular viability and optimizing measurement conditions. Most of them focused on static or slowly evolving samples, representing the changing chemical composition over the cell cycle and/or in response to external stimuli, and provided relatively large absorbance changes, of 0.01- 0.1 au over the course of a few hours, which are easily accessible by IR microscopy. (Jamin 1998, Holman 2000a, Holman 2000b, Holman 2002, Heraud 2005, Moss 2005, Zhao 2009, Kuimova 2009, Nasse 2009, Marcsisin 2010, Tobin 2010) In several cases cellular spectra were analysed with the same multivariate classification algorithms in use with dried or fixed cell samples, with a view towards diagnostic applications. The following decade saw the extension of these measurements to more challenging cellular targets, from a biochemical perspective, including the spectroscopy of specific proteins and the measurement of formation and consumption of small molecules, as detailed in the following sections.

*Measurement of specific proteins in single living and functional cells.*

Part of the attractiveness of live cell experiments is the possibility to extend the measurements traditionally performed *in vitro*, on purified proteins or other biomolecules, to *in vivo* or *ex vivo* conditions. Any such studies must overcome two challenges, the low concentration of the protein of interest in the cellular environment and the complexity of said environment, requiring both high sensitivity and high selectivity. Total protein content in a mammalian cell is of the order of ~0.3-0.9 M peptide bond concentration in a single cell, and is easily measurable by IR microscopes. However, the concentration of individual proteins varies over orders of magnitude. The most abundant ones, including some extracellular secreted proteins, such as collagen and mucin, are locally present at high concentration, and can be measured selectively by IR microscopy. Deposits of amyloid proteins are also accessible to direct investigation. (Choo 1996, Miller 2002) In specialized cells and cellular compartments, local concentrations of some specific intracellular proteins can be within easy reach. Two notable examples are haemoglobin in red blood cells (Dong 1988) and rhodopsin in the outer segment of retinal rod cells (Quaroni 2008), which are present at mM concentration. However, in more general cases, abundant intracellular proteins are in the 10-100 μM range. For such concentrations, measurement of the global absorption of the peptide backbone is accessible, but measurement of individual amino acid residues and cofactors is limited or precluded. Beyond these cases, the concentration of most proteins is well below the detection limit of FTIR experiments, preventing observation even in bulk cellular samples. Single cell measurements further exacerbate these constraints, and few reports exist of IR spectromicroscopy of individual proteins in single living cells.

Rhodopsin in single rod cells is one such case. Rod cells are the photoreceptors responsible for vision under dim light conditions and perform their function by relying on a highly specialized cellular compartment, the rod outer segment (ROS). Amphibian ROS are about 30-50 μm in length and 4-6 μm in diameter, providing a good match to mid-IR spatial resolution. The proteome of the ROS is dominated by the photon-sensing protein rhodopsin, a transmembrane protein folded into a seven-helix bundle. The subcellular environment is structured in a stack of disk-shaped membranes, with the alfa helices of rhodopsin aligned perpendicular to the disk surface and parallel to the longitudinal axis of the ROS (Figure 4). The anisotropic structure permits linear dichroism measurements on single ROS, and the polarized absorption of the Amide I band could be used to quantify the average angle of the alfa helices relative to the cell axis. For this experiment, the capability of measuring a single ROS is a distinct advantage, because it avoids the spread of angular orientations that is unavoidable in the measurement of ensembles of cells or membrane fragments, and provides angular values close to the one obtained from X-ray diffraction on protein crystals (pdb 1f88 entry; Figure 4B). (Quaroni 2008) Single ROS measurements with synchrotron radiation were used to follow the response of the system to illumination. The time resolution, limited to a few seconds, excluded most of the rhodopsin photocycle from investigation. Nonetheless, it was possible to identify slower spectroscopic changes, attributed to the final stages of retinal release and the associated reorganization of the whole outer segment (Figure 4C). Fitting of the traces in Figure 4C provides time constants for the respective transformations. (Quaroni 2011b) At a higher level of complexity, time resolved measurements in the micro-ATR configuration were performed on single retinas, reporting the spatial reorganization of the lipid architecture of the cells that accompany light exposure. Some of the spectral changes were also tentatively assigned to the redistribution of the G-protein transducin, the second most abundant protein in rod cells. (Massaro 2008) If confirmed, this experiment would provide another example of an *ex vivo* measurement of a specific protein.

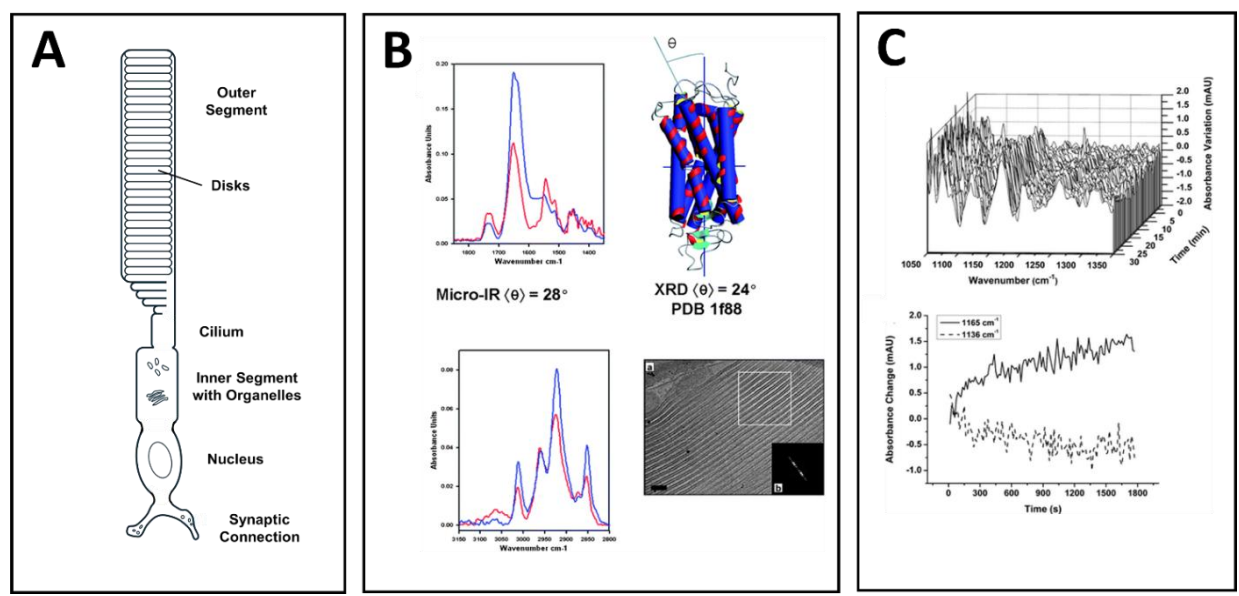

*Figure 4. IR microscopy of single retinal rod cells.* A. Schematic representation of a rod photoreceptor. B. Linear dichroism of the outer segment of an amphibian rod cell. Measurements of linearly polarized spectra provide the average angle of protein alfa helices relative to the rod axis (upper spectrum: perpendicular to the axis; lower spectrum: parallel to the axis), in good agreement with the value extracted from the crystal structure of rhodopsin. The result is consistent with the regular orientation of outer segment disks from electron microscopy (bottom right panel). C. Time resolved spectral measurements on the outer segment following light exposure (upper panel) and single wavelength kinetic traces extracted from the spectra (lower panel), corresponding to molecular changes associated with the slowest, terminal stages of the photocycle. Panel B modified from Quaroni 2008, Quaroni 2011a, and Nickell 2007. Panel C reproduced with permission from Quaroni 2011b.

Most proteins of interest do not enjoy such favourable conditions as rhodopsin and are not measurable at their natural concentration. One way to bypass such limitation has been heterologous overexpression of eukaryotic proteins in prokaryotes, to achieve measurable concentrations. (Mészáros 2020; Goett-Zink 2020) While the cytoplasmic environment is not the same as that of the native cellular system and the use of prokaryotic cells limits the possibility of measuring single cells, the approach is a useful proof of concept for a strategy to investigate proteins *in vivo*, inside functional cells. The next desirable milestone is homologous overexpression in eukaryotic cells, which would allow single cell investigations of the protein in interaction with its natural partners.

*Measurement of small molecules in cells and tissue.*

Like the study of individual protein activity, the detection and quantification of small molecule metabolites is a motivation for the development of IR microscopy of live cells. Small molecules provide critical information about cellular biochemistry, but their content is easily perturbed, and they are degraded or lost when cells are dried or undergo fixation. The concentration of specific metabolites depends on cell type and conditions and can fluctuate widely during the life of the cell. Detection limits vary with metabolite and conditions, and depend on the absorption coefficient of specific molecules, sample thickness, instrumental performance and measurement time. Given the current performance of FTIR instruments, detection limits for small molecules can reach $10^{-1}$ - $10^{-2}$ mM when measuring strong bands, and 1.0 – 0.1 mM for IR microscopes.

To form an idea about detectability, it is helpful to consider the concentrations of the more abundant ones. Some soluble small metabolites, such as lactate, the acids of the TCA cycle or the adenine nucleotides, can reach mM concentrations, and are generally accessible to IR microscopy experiments. (Quaroni 2016, Quaroni 2020a) Molecules at lower concentrations may still be detectable under appropriate conditions and in more demanding measurements, depending on their absorption coefficient.

In contrast to soluble metabolites, lipids also have a structural role and are mostly bound up in membranes. Given the abundance of eukaryotic intracellular membranes, lipids account for more than 10% of the dry mass of mammalian cells, and their overall contribution to cellular IR absorption is substantial.

As a guideline to the spectral contribution of small metabolites, we can assess the combined abundance of molecules with the same functional group, to obtain a cellular "functional group concentration". Considering soluble metabolites, the concentration of some common functional groups, such as carboxylic acids, amides and phosphate esters, can reach the 10's of mM, and is appreciable in IR spectra. These values also account for metabolites that are individually below the detection limit, but the total contributions of which can easily reach the millimolar range.

A similar exercise can be carried out with the lipid content. The concentration of phosphate and carboxylic esters associated to phospholipids is hundreds of mM in fully hydrated cells. The concentration of methylene groups can be as high as 1 - 2 M in hydrated cells, which accounts for the appearance of the characteristic sharp doublet at ~2855 cm$^{-1}$ and ~2920 cm$^{-1}$ in the spectra of intact cells (Figure 1). Owing to the intensity of such bands, gross changes in the cellular metabolism of phospholipids and fatty acids are easily identified in live cell measurements, without further analysis. (e.g. Heraud 2005; Marcsisin 2010; Vannocci 2018)

The concentration of peptide bonds from small metabolites and lipids corresponds to a few percentage points of the value for main chain peptide bonds of 300 – 900 mM (see previous section). However, the concentration of other functional groups with an overlapping spectral range, such as carboxylate and ammonium groups, can jointly surpass 100 mM, giving an appreciable contribution.

Another noteworthy term of comparison is the contribution from nucleic acids. The concentration of nucleotide monomers in DNA is in the mM range in a human cell, while that of RNA is three to five times larger, giving an overall concentration of nucleotide monomers in nucleic acids between 10 mM and 20 mM. This is smaller than the concentration of phosphate ester groups associated to small metabolites and lipids. Despite this caveat, many publications routinely report the observation of DNA and RNA absorption in dried, fixed or live cells and tissue (including past work by the present author and several articles referenced in this work) in the phosphate absorption region, but without any systematic band assignments. In view of the concentration of nucleic acids, existing conclusions in the literature should be reassessed and confirmed.

The general conclusion of this analysis is that the contribution of small metabolites to the IR spectra of cells and tissue is appreciable, when compared with that from the main classes of biopolymers. The conclusion is consistent with the proposal by Ellis and Goodacre (Ellis 2006) that the spectral contribution of metabolic activity allows cellular discrimination and classification for diagnostic purposes. On the reverse side of the coin, the spectral contributions from the bulk metabolome overlap with those of any specific molecules that an investigator may be interested in, and can be a severe interference in molecular identification and band assignment.

These considerations highlight the potential impact of IR microscopy in metabolomic applications. The sensitivity is much lower than that of mass spectrometry, and somewhat lower than that of NMR, and only one or two dozen or so molecules can be detected. Nonetheless, the technique allows the measurement of multiple dynamic processes, down to a single cell, expanding the view of metabolomics to a different perspective, that of spatially resolved reactions.

*Small molecule turnover in single living and functional cells*

The direct measurement of molecular turnover, whether of metabolites or of exogenous molecules, is the mainstay of a biochemical analysis and relies heavily on the use of difference spectra. At the single cell level, time-resolved difference spectroscopy can be extended to imaging measurements, by collecting hypercubes of difference spectra. In the resulting IR images, contrast is determined by the local change in absorbance at a specific wavelength over time and can be used to visualize the progressive formation of concentration gradients of small molecules in the proximity of a cell. (Quaroni 2014c; Quaroni 2016)

Measuring reactions in a single cell places combined demands on sensitivity and on selectivity, because of low concentrations and sample complexity. Spectral variations need to be detected, and also need to be assigned to the correct molecule, as a precondition for any quantitative measurements. (Baker 2016, comments by L. Quaroni therein) If the molecule of interest absorbs strongly in a spectrally clean region, the bands can be used directly for identification and quantitation. These cases are exceptions rather than the rule, although they are visibly represented in the literature, because of the relative ease of the measurements. A point in case is provided by di-nuclear and tri-nuclear molecules with multiple bonds, which have strong absorption bands due to their stretching modes (ν) in the frequency range between 1800 and 2500 cm$^{-1}$. This range has few other strong absorptions, allowing their unhindered observation. Molecules in this group include several endogenous and exogenous compounds of biological relevance, such as carbon and nitrogen oxides, cyanide, azide and their derivatives. Carbon monoxide is one of the most relevant ones in a biological context, in part because of its role as an endogenous signalling molecule. Its metal complexes can be easily measured using stretching mode bands, which also provide valuable structural information because of their sensitivity to geometry and molecular environment. These properties have been extensively exploited for the investigation of metal carbonyl complexes in biological systems, some of which are of pharmacological relevance and can be characterized in terms of activity and toxicity. (Butler 2012; Quaroni 2014a, Santoro 2015) The sensitivity of the measurement has allowed IR microscopy studies of their uptake, distribution and dynamics in single live cells and tissues. (Zobi 2013, Santoro 2015, Clède 2020) Figure 5 shows an example where the stretching bands of a facial tricarbonyl complex (ReC$_{12}$, Figure 5A) are used to quantify the kinetics of cellular uptake, (Clède 2020) in the form of single band kinetics (Figure 5B) or spectral kinetics (Figure 5C). The sharp rises and dips in the traces of Figure 5B are attributed to the movement of the cell. Fitting the traces from single cell experiments to a Hill model provides support for a mechanism of active cellular uptake.

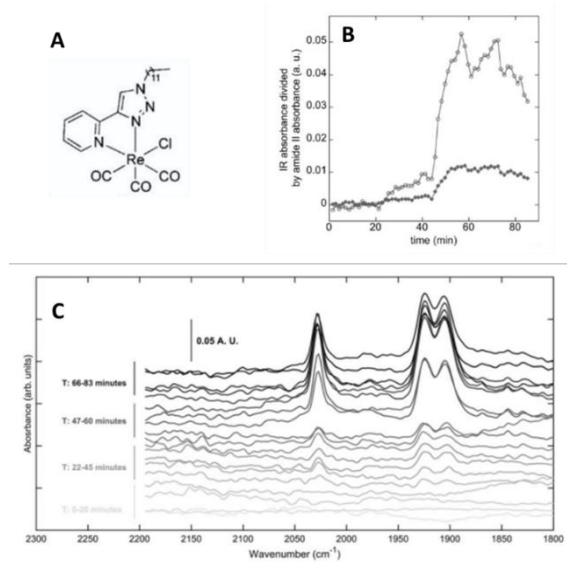

*Figure 4. IR microscopy and spectroscopy of uptake of a Re(CO)$_3$ complex by a single cell. A: Structure of the fac tri-carbonyl complex ReC$_{12}$. The complex has a long alky chain to facilitate uptake. B: Time course of the carbonyl bands relative to the Amide II band following*

addition of 25 μM ReC$_{12}$ to the cell culture. Gray trace: area of bands at 1940–1890 cm$^{-1}$; black trace: area of band at 2035–2020 cm$^{-1}$. C: Changes in the carbonyl region of the cellular spectrum during uptake. Modified from Clède 2020.

Among other molecules that absorb in the 1800 and 2500 cm$^{-1}$ interval, carbon dioxide deserves particular attention, being the catabolic end product of organic molecules. The strong absorption band of aqueous CO$_2$ (CO$_{2(aq)}$) at 2343 cm$^{-1}$ ($\nu_{asCO2}$) allows its quantification. (Falk 1992) Aqueous concentrations of CO$_{2(aq)}$ at normal atmospheric composition are in the 10 μM range, corresponding to the solubility. However, common cell culturing conditions use approx. 5% environmental CO$_2$, leading to mM concentrations in the medium, which are easily measured. Dissolved CO$_2$ (CO$_{2(aq)}$) rapidly hydrates to form carbonic acid, which equilibrates with bicarbonate in the medium. The presence of the CO$_2$/HCO$_3^-$ acid-base pair in cell culture media provides the opportunity to rely on their relative abundance for optical monitoring of pH variations. Addition of H$^+$ ions to the medium decreases [HCO$_3^-$] and increases [CO$_{2(aq)}$], which can be related to the corresponding change in pH. (Equation 1)

$$pH = 6.04 - \log([CO_2]/[HCO_3^-]) \qquad \textbf{Equation 1}$$

Equation 1 has been used to monitor the drop in pH associated to anaerobic glycolysis in cell cultures and at single cells, (Quaroni 2014b, Vannocci 2021) and to estimate local acidification caused by the photolysis of caged proton-releasing molecules. (Carbone 2013) As an example, Figure 6 shows the spectral changes that accompany glycolytic turnover of single adherent cells from the *HEK-cFXN* cell line (a cell line derived from HEK293). Lactate formation is simultaneously detected in the same measurement. (Vannocci 2021)

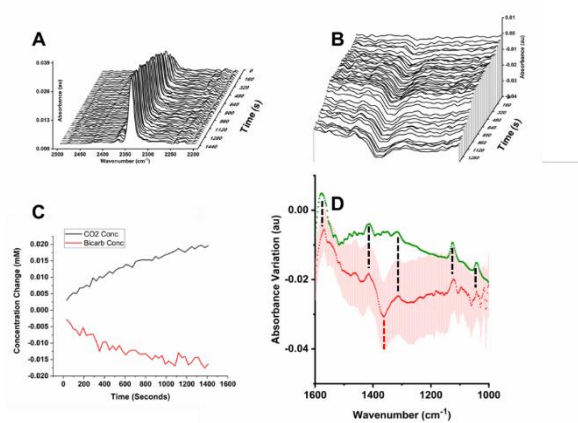

*Figure 6. Quantitative Measurements of Acidification and Lactic Fermentation of Single HEK-cFXN Cells by IR spectromicroscopy. The figure shows difference spectra at selected spectral intervals. **A.** Increase of the absorption band from aqueous CO$_2$ formed from acidification of the cellular environment. **B.** Parallel decrease of the absorption band from aqueous HCO$_3^-$ caused by acidification. **C.** Time course of the changes in concentration of aqueous CO$_2$ and aqueous HCO$_3^-$ extracted from*

*difference spectra.* ***D.*** *Spread of 3 difference spectra measured after 25 min on single cells in the 1000 – 1600 cm$^{-1}$ interval (red trace: average; pink area: σ). The dashed red line marks the position of the decreasing HCO$_3^-$ absorption band. The green trace shows by comparison the spectrum of a 10 mM sodium lactate solution. The dashed black lines mark the position of absorption bands from lactate as they appear in cellular difference spectra. Modified from Vannocci 2021.*

The measurements in Figure 6 also show an example where lactate, a small molecule metabolite, is detected. Lactate is relatively abundant in mammalian eukaryotic cells, typically in the mM range. In contrast to the small molecules in the previous sections, and similarly to most metabolites, its main absorption bands fall in crowded spectral regions, and its identification is contingent on the detection and assignment of multiple bands (see following sections). Lactate is formed as the end product of glycolysis and fermentation under hypoxic conditions, before being exported to the extracellular environment. No other intermediates of glycolysis were detected in these measurements, presumably because homeostasis maintains stable concentrations. However, weak spectral changes assigned to some of these intermediates were detected in multicellular measurements of A549 cells by the use of 2D-COS analysis. The outcome was attributed to the progressive onset of hypoxia in the sample, which delays the establishment of steady-state conditions. (Quaroni 2014b) Similarly, ethanol production by *Chlamydomonas reinhardtii* (Goff 2009), which is the end product of fermentation, is easily quantified. In contrast, most intermediates of metabolic pathways are present at measurable concentration but are difficult to observed in time-resolved difference spectra because their concentration changes slowly over the measurement. Detection of these intermediates necessitates perturbation of steady-state conditions and requires careful experimental design.

*Metabolic Water*

Water is the most abundant cellular metabolite, a fact which is often overlooked in metabolic analysis. Water molecules are released during the synthesis of biological macromolecules and some small molecules. Water is also formed from the terminal reduction of dioxygen. Conversely, water molecules are consumed during the corresponding inverse reactions. The strong absorption of IR radiation by water makes IR spectroscopy an ideal technique to follow their turnover, although the spectral contribution of metabolic water must be separated from that of bulk water. One approach is to rely on the effect of metabolic activity on the isotopic composition of intracellular water. (Kreuzer-Martin 2005) Prior to the measurement, cells are submerged in a medium prepared with isotopically marked water molecules, which shifts the absorption bands of the bulk solvent. Water formed from the initial metabolic turnover still retains the natural isotopic composition and can be selectively identified. Proof of concept experiments on fibroblasts in $^2$H$_2$O medium have shown that $^1$H$_2$O production can be monitored indirectly, via the isotopically mixed water molecule $^1$HO$^2$H, which forms by hydrogen exchange with the bulk $^2$H$_2$O environment (Figure 7). The experiments highlighted the close connection between metabolic activity and water production: *de novo* production of water molecules was observed only for fibroblast cultures in the growth phase, during which extensive synthesis of macromolecules takes place. In contrast, negligible water production was observed for stationary cultures (Figure 7A-D). (Kreuzer 2012) Single cell IR imaging shows the formation of a concentration gradient of $^1HO^2H$ in the proximity of cells surrounded by $^2H_2O$ medium (Figure 7E-F), thus excluding that isotopic enrichment could be caused by water molecules absorbed from the atmosphere. (Quaroni 2014)

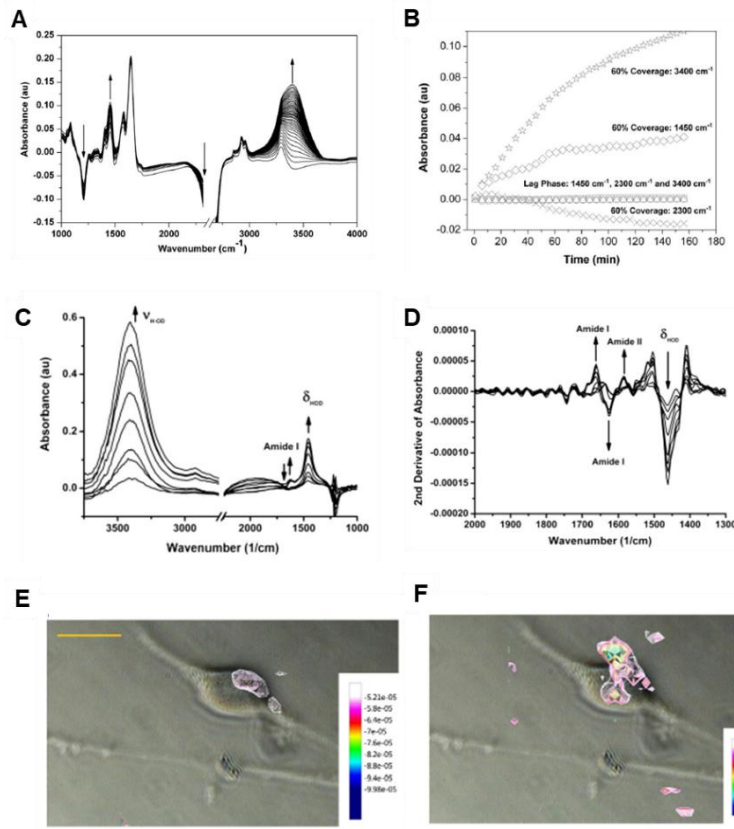

*Figure 7. IR spectroscopy and microscopy of metabolic water turnover in fibroblast cells. A: Time-resolved difference spectra of a culture of Rat-1 fibroblasts in $D_2O$ medium recorded over 360 min. The arrows show the bands of the water molecules with mixed isotopic composition, HOD, that form over time. B: Evolution of HOD bands over 180 min in fibroblast cultures either growing (60% coverage) or in the lag phase. Lag phase cells do not show any measurable formation of HOD. C: Time resolved difference spectra recorded over 160 min at a single NIH 3T3 Swiss albino fibroblast cell, showing local formation of HOD. The arrows indicate the direction of change of bands responsive to D substitution. D: 2nd derivative of difference spectra in C below 2000 cm$^{-1}$. E: IR image of the gradient of HOD concentration in the proximity of a single cell, collected soon after exposing the cell to $D_2O$ medium. Image constructed from the amplitude of the 2nd derivative of the $\delta_{HOD}$ band at 1450 cm$^{-1}$ (More negative values correspond to higher absorption). F: IR image of gradient of HOD concentration 200 min after exposing the cell to $D_2O$ medium, showing the increasing local concentration of HOD (Note that the scale is different from the one in E). Modified from Kreuzer 2012 and Quaroni 2014c.*

*Methods for Molecular Identification*

In as complex a sample as a living cell, detection of a molecule is contingent on the possibility to identify its spectral bands among the forest of contributions from other components. Except for the most abundant molecules, assignments that rely simply on the observation of individual bands, are often inadequate. Chemical identification via *fingerprinting* is often touted as a hallmark of IR spectroscopy and is the basis of untargeted molecular identification. However, fingerprinting requires resolution of the extended band pattern of a specific molecule, in terms of both position and intensity. Because the absorption bands of most detectable metabolites fall within crowded spectral intervals, to reliably identify the bands of specific molecules, methods of spectral assignment are required. The following sections will discuss two of them, isotopic labelling and correlation analysis.

- *Isotopic Labelling*

Isotopic labelling is an established practice in vibrational spectroscopy. Replacement of an atom with a heavier isotope downshifts the frequency of absorption bands that involve its displacement of the labelled atom. (Wilson 1980) The practice is particularly effective when complemented by theoretical analysis, because isotopic shifts can be calculated with better accuracy than absolute vibrational frequencies, even with basic levels of theory and, in the case of smaller molecules, with simple analytical expressions. Over the past decades, isotopic shifts have been considered a gold standard for band assignment. (Barth 2003, Berthomieu 2009)

The application of isotopic labels is not limited to mode assignment. In complex samples, isotopic substitution allows isolating the absorption bands of labelled molecules from those of similar but unlabelled ones. The use of isotopic tracers is also a standard method in metabolic analysis, to follow the fate of molecules and constituent atoms through metabolic pathways and through space, at the cellular, tissue and whole organism level. Not least, tracing the incorporation of isotopic labels in reaction products and reaction intermediates is part of the toolbox for structure-function studies of biochemical reactions, notably enzymatic catalysis. In IR microscopy of living cells, these applications can be simultaneously incorporated into a single experiment, providing a wealth of information. Despite the multiple advantages, isotopic labelling experiments come with a financial cost, corresponding to the expense and synthetic effort of preparing the labelled molecule.

- *Correlation Analysis*

While isotopic labelling is an established practice in vibrational spectroscopy, correlation analysis is a relatively recent introduction. The use of 2D correlation spectroscopy (2D-COS) provides a tool to facilitate band assignment and molecular identification without the need for target selection. It was initially used to identify the spectroscopic contributions in the time resolved micro ATR spectra of retinas following illumination (Massaro 2008) and was later developed into a systematic protocol for band assignment (Quaroni 2011b, Quaroni 2014b). 2D-COS was originally introduced by Noda as a procedure to extract information from sequential sets of spectroscopic data using the two-dimensional Fourier-transform (2D-FT) of their correlation function. (Noda 2004, Noda 2009) The parameter that defines the sequence can be time or any evolving quantity, including temperature, pressure, and electrochemical potential, among others. The outcome of the 2D-FT is a 2D complex function that represents the phase relationship between spectral changes. The real part of the function, termed the synchronous plot, shows the correlation between spectral bands that change with a phase relationship of +/- π, that is in phase or with opposite phase. The imaginary part of the function, termed the asynchronous plot, shows the correlation between bands that change sequentially, with any phase relationship different from +/- π.

The use of 2D-COS for band assignment takes advantage of the property that bands arising from the same molecule change with perfect synchrony and show strong correlation peaks in the synchronous plot but no peaks in the asynchronous plot. Comparison of synchronous and asynchronous plots can cluster bands into groups that satisfy this condition. Each of these groups corresponds to the bands of a single species or of multiple species that evolve in synchrony. As an example, the products of a reaction A → B + C would give a single group of bands that form in synchrony. 2D-COS analysis can also be applied to IR maps and images by using the sequence of pixels as the variable parameter. Because the spatial distribution of bands tracks the spatial distribution of the compound they arise from, 2D-COS analysis of spatial spectral sequences can also be used for the purpose of band assignment. (Lasch 2019; Quaroni 2020b)

Validation of 2D-COS for molecular identification was provided in the measurement of the light response of a retinal rod cell, where it was used to identify contributions from different molecular components (Figure 7 A and B). (Quaroni 2011b) Nearly all measured bands could be assigned, leading to the resolution of spectral changes into three different molecular contributions, and interpretation of associated kinetic processes. In separate work, the capability of synchronous plots to extract weaker spectral contributions has also been exploited for the recovery of bands in noisy spectra. (Quaroni 2010, Zobi 2013) The workflow leading from spectromicroscopy measurements to molecular identification and kinetic analysis has been introduced under the umbrella denomination of Correlated Cellular Spectro-Microscopy (CSM), and has been proposed as a general protocol for the analysis of metabolism in living cells. (Quaroni 2014b)

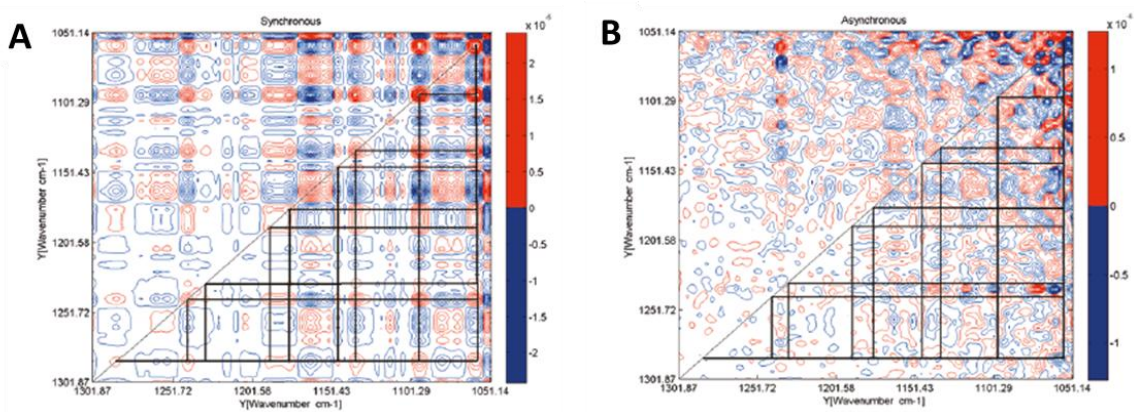

*Figure 7. Band assignment using 2D-COS. A: Synchronous Plot calculated from the time resolved spectra of a rod cell in Figure 4. The lines connect a group of peaks that change in phase. B: Asynchronous Plot calculated from the time resolved spectra of Figure 4. The lines connect the same*

*positions as in Panel A, showing a lack of asynchronous peaks, corresponding to lack of out-of-phase changes. It is concluded that the peaks evolve with perfect synchrony and must arise from the same molecule. Modified from Quaroni 2011b.*

## Conclusions & Future Perspectives

The quantitative application of IR microscopy to living cells has reached a level of development that can impact our understanding of cellular biology. At the core of this promise is the capability of quantitatively and simultaneously monitoring a variety of biochemical processes in single cells in real time without *a priori* selection of a target. The implementation of imaging configurations expands these capabilities by providing a two-dimensional distribution of molecular absorbers in the sample and the visualization of cellular chemistry in space and time. The past twenty years have seen the laying of methodological and technical foundations and the implementation of basic experiments. The immediate future will see the application of live cell IR microscopy to address specific issues in cellular biology, paramount among them the quantitative description of cellular metabolism. Historically, developments in microscopy techniques have expanded our understanding of biology, from the introduction of the first optical microscopes, which revealed cells and microbial life, to the invention of the electron microscope, which brought to our eyes the nanoscale architecture of cellular space, membranes, proteins and nucleic acids. The promise of IR microscopy that is unfolding is to complement existing views with a quantitative description of reaction turnover, metabolic activity and cellular biochemistry, in single cells and in their environment.

## Funding

The preparation of this manuscript was supported by an OPUS 25 grant from the National Science Centre Poland under agreement 2023/49/B/NZ1/00768.

## Acknowledgments

The author is grateful to Annalisa Pastore, Tommaso Vannocci, Theodora Zlateva, and Fabio Zobi for reading the manuscript and for helpful comments.

## References

Aas, E. *"Refractive Index of Phytoplankton Derived from its Metabolite Composition"; J. Plankton Res.* **1996**, 18, 12, 2223-2249. DOI: 10.1093/plankt/18.12.2223.

Arrondo, J.L.R., A. Muga, J. Castresana, F. M. Goñi *"Quantitative Studies of the Structure of Proteins in Solution by Fourier-Transform Infrared Spectroscopy"; Prog. Biophys. Mol. Biol.* **1993**, 59, 1, 23-56. DOI: 10.1016/0079-6107(93)90006-6.

Baker, M.J., *et al.* *"Single Cell Analysis/Data Handling: General Discussion"; Faraday Discuss.* **2016**, 187, 299-327. DOI: 10.1039/c6fd90012g

Barth, A. *"The Infrared Absorption of Amino Acid Side Chains"*; Prog. Biophys. Mol. Biol. **2000**, 74, 141–173. DOI: 10.1016/s0079-6107(00)00021-3

Barth, A., C. Zscherp *"What Vibrations Tell Us about Proteins"*; Q. Rev. Biophys. **2002**, 35, 4, pp. 369–430. DOI: 10.1017/S0033583502003815.

Brubach, J.-B., A. Mermet, A. Filabozzi, A. Gerschel, D. Lairez, M. P. Krafft, P. Roy *"Dependence of Water Dynamics upon Confinement Size"*; The Journal of Physical Chemistry B **2001** 105 (2), 430-435 DOI: 10.1021/jp002983s

Butler, I.S., R.P. Kengne-Momo, G. Jaouen, C. Policar, A. Vessières *"Recent Analytical Applications of Molecular Spectroscopy in Bioorganometallic Chemistry—Part I: Metal Carbonyls"*; Appl. Spectrosc. Rev. **2012** 47:7, 531-549, DOI:10.1080/05704928.2012.673189

Carbone, M., T. Zlateva, L. Quaroni *"Monitoring and Manipulation of the pH of single Cells Using Infrared Spectromicroscopy and a Molecular Switch"* Biochim. Biophys. Acta (BBA) - General Subjects **2013**, 1830, 4, 2989-2993. DOI: 10.1016/j.bbagen.2012.12.022.

Chan, K.L.A., P.L.V. Fale, A. Atharawi, K. Wehbe, G. Cinque *"Subcellular Mapping of Living Cells via Synchrotron MicroFTIR and ZnS Hemispheres"*; Anal. Bioanal. Chem. **2018,** 410, 6477–6487. DOI: 10.1007/s00216-018-1245-x

Chan, K.L.A , S.G. Kazarian *"Attenuated Total Reflection Fourier-Transform Infrared (ATR-FTIR) Imaging of Tissues and Live Cells"*; Chem. Soc. Rev. **2016**, 45, 1850. DOI: 10.1039/c5cs00515a

Choo, L.P., D.L. Wetzel, W.C. Halliday, M. Jackson, S.M. LeVine, H.H. Mantsch *"In Situ Characterization of Beta-Amyloid in Alzheimer's Diseased Tissue by Synchrotron Fourier Transform Infrared Microspectroscopy"*; Biophys. J. **1996,** 71, Issue 4, 1672-1679, DOI: 10.1016/S0006-3495(96)79411-0.

Clède, S., C. Sandt, P. Dumas, C. Policar *"Monitoring the Kinetics of the Cellular Uptake of a Metal Carbonyl Conjugated with a Lipidic Moiety in Living Cells Using Synchrotron Infrared Spectromicroscopy"*; Appl. Spectrosc. **2020**, 74(1) 63–71 DOI: 10.1177/0003702819877260

Diem, M., S. Boydston-White, L. Chiriboga *"Infrared Spectroscopy of Cells and Tissues: Shining Light onto a Novel Subject."*; Appl. Spectrosc. **1999**, 53, 4, 148A-161A DOI:10.1366/0003702991946712

Dong, A., R.G. Messerschmidt, J.A. Reffner, W.S. Caughey, *"Infrared spectroscopy of a single cell — the human erythrocyte"*; Biochem. Biophys. Res. Comm., **1988**, 156, 2 752-756 https://doi.org/10.1016/S0006-291X(88)80907-0.

Ebbinghaus, S., S.J. Kim, M. Heyden, X. Yu, U. Heugen, M. Gruebele, D.M. Leitner, M. Havenith *"An Extended Dynamical Hydration Shell Around Proteins"*; Proc. Natl. Acad. Sci. **2007**, 104, 52, 20749-20752. DOI: 10.1073/pnas.0709207104.

Ellis, D.I.; R. Goodacre *"Metabolic Fingerprinting in Disease Diagnosis: Biomedical Applications of Infrared and Raman Spectroscopy"*; Analyst **2006**, 131, 875-885. DOI: 10.1039/b602376m.

Falk, M.; A.G. Miller  *"Infrared Spectrum of Carbon Dioxide in Aqueous Solution"*; Vib. Spectrosc. **1992**, 4, 1, 105-108, DOI: 10.1016/0924-2031(92)87018-B.

Fabian, H., W. Mäntele "*Infrared Spectroscopy of Proteins.*" in *Handbook of Vibrational Spectroscopy* J. Chalmers, P. Griffiths (Eds.). John Wiley & Son, Chichester, **2002** 5, 3399-3425. ISBN-10 0471988472; ISBN-13 978-0471988472

Fitzpatrick, J., J.A. Reffner *"Macro and Micro Internal Reflection Accessories"* in *Handbook of Vibrational Spectroscopy* J. Chalmers, P. Griffiths (Eds.)  John Wiley & Son, Chichester, **2002** 2, 1103-1116. ISBN-10 0471988472; ISBN-13 978-0471988472

Goett-Zink L., J.L. Klocke JL, L.A.K. Bögeholz, T. Kottke *"In-cell Infrared Difference Spectroscopy of LOV Photoreceptors Reveals Structural Responses to Light Altered in Living Cells."* J. Biol. Chem. **2020**; 295(33):11729-11741. DOI: 10.1074/jbc.RA120.013091.

Goff, K.L., L. Quaroni, L., K.E. Wilson *"Measurement of Metabolite Formation in Single Living Cells of Chlamydomonas reinhardtii Using Synchrotron Fourier-Transform Infrared Spectromicroscopy"*; Analyst **2009**,134, 2216-2219 DOI: 10.1039/b915810c

Heraud, P.,  B.R. Wood, M.J. Tobin, J. Beardall, D. McNaughton *"Mapping of Nutrient-Induced Biochemical Changes in Living Algal Cells Using Synchrotron Infrared Microspectroscopy"*; FEMS Microbiol. Lett. **2005**, 249, 2, 219–225, DOI: 10.1016/j.femsle.2005.06.021

Holman, H-Y. N., M. C. Martin, E. A. Blakely, K. Bjornstad, W. R. McKinney *"IR Spectroscopic Characteristics of Cell Cycle and Cell Death Probed by Synchrotron Radiation Based Fourier Transform IR Spectromicroscopy"*; Biopolym. **2000a**, 57, 329–335. DOI: 10.1002/1097-0282(2000)57:6<329::AID-BIP20>3.0.CO;2-2

Holman, H-Y. N.,  R. Goth-Goldstein, M. C. Martin, M. L. Russell, W. R. McKinney *"Low-Dose Responses to 2,3,7,8-Tetrachlorodibenzo-p-dioxin in Single Living Human Cells Measured by Synchrotron Infrared Spectromicroscopy"*;  Environ. Sci. Technol. **2000b**, 34, 2513–2517. DOI: 10.1021/es991430w

Holman, H-Y. N., K. A. Bjornstad, M. P. McNamara, M. C. Martin, W. R. McKinney, E. A. Blakely *"Synchrotron Infrared Spectromicroscopy as a Novel Bioanalytical Microprobe for Individual Living Cells: Cytotoxicity Considerations"* J. Biomed. Opt. **2002**, 7, 417–424. DOI: 10.1117/1.1485299.

Jamin N., P. Dumas, J. Moncuit, W.H. Fridman, J.L. Teillaud, G.L. Carr, G.P. Williams *"Highly Resolved Chemical Imaging of Living Cells by Using Synchrotron Infrared Microspectrometry."* Proc. Natl. Acad. Sci. USA. **1998** Apr 28;95(9):4837-40. DOI: 10.1073/pnas.95.9.4837.

Kreuzer-Martin, H.W.; J.R. Ehleringer, E.L. Hegg *"Oxygen Isotopes Indicate Most Intracellular Water in Log-Phase Escherichia coli is Derived from Metabolism"*; Proc. Natl. Acad. Sci. U.S.A. **2005**, 102 (48) 17337-17341, DOI: 10.1073/pnas.0506531102.

Kreuzer , H.W., L. Quaroni , D. W. Podlesak, T. Zlateva, N. Bollinger, A. McAllister, M.J. Lott, E.L. Hegg *"Detection of Metabolic Fluxes of O and H Atoms into Intracellular Water in Mammalian Cells"*; PLoS One **2012** 7(7): e39685, DOI: 10.1371/journal.pone.0039685.

Kuimova, M. K, K. L.A. Chan, S.G. Kazarian *"Chemical Imaging of Live Cancer Cells in the Natural Aqueous Environment."*; Appl. Spectrosc. **2009,** 63, 2, 164-171. DOI: 10.1366/000370209787391969.

Lasch P, L. Chiriboga, H. Yee, M. Diem *"Infrared Spectroscopy of Human Cells and Tissue: Detection of Disease"*; Technol. Cancer Res. Treat. **2002**; 1(1):1-7. DOI: 10.1177/153303460200100101

Lasch, P., I. Noda *"Two-Dimensional Correlation Spectroscopy (2D-COS) for Analysis of Spatially Resolved Vibrational Spectra"*; Appl. Spectrosc. **2019**, 73, 359-379. DOI: 10.1016/j.molstruc.2020.128068

Lewis, E.N., P. J. Treado, R.C. Reeder, G.M. Story, A.E. Dowrey, C. Marcott, I.W. Levin *"Fourier Transform Spectroscopic Imaging Using an Infrared Focal-Plane Array Detector"*; Anal. Chem. **1995** 67 (19), 3377-3381. DOI: 10.1021/ac00115a003

Marcsisin, E.J., C.M. Uttero, M. Miljković, M. Diem *"Infrared Microspectroscopy of Live Cells in Aqueous Media"*; Analyst, **2010**, 135, 3227–3232. DOI: 10.1039/c0an00548g

Martin, M.C., C. Dabat-Blondeau, M. Unger, J. Sedlmair, D. Y. Parkinson, H. A. Bechtel, B. Illman, J. M. Castro, M. Keiluweit, D. Buschke, B. Ogle, M. J. Nasse, C. J. Hirschmugl *"3D Spectral Imaging with Synchrotron Fourier Transform Infrared Spectro-Microtomography"*; Nat. Methods **2013**, 10, 861–864. DOI: 10.1038/nmeth.2596

Massaro, S., T. Zlateva, V. Torre, L. Quaroni *"Detection of Molecular Processes in the Intact Retina by ATR-FTIR Spectromicroscopy"*. Anal. Bioanal. Chem. **2008**, 390, 317–322. DOI: 10.1007/s00216-007-1710-4

Mattson, E.C., M. Unger, S. Clède, F. Lambert, C. Policar, A. Imtiaz, R. D'Souza, C.J. Hirschmugl *"Toward Optimal Spatial and Spectral Quality in Widefield Infrared Spectromicroscopy of IR Labelled Single Cells"*; Analyst **2013**, 138, 5610

Mészáros, L.S., P. Ceccaldi, M. Lorenzi, H.J. Redman, E. Pfitzner, J. Heberle, M. Senger, S.T. Stripp, G. Berggren *"Spectroscopic Investigations Under Whole-Cell Conditions Provide New Insight into the Metal Hydride Chemistry of [FeFe]-Hydrogenase"*; Chem. Sci. **2020**, 11, 4608–4617 DOI: 10.1039/d0sc00512f

Miller, L.M., P. Dumas, N. Jamin, J.-L. Teillaud, J. Miklossy, L. Forro *"Combining IR Spectroscopy with Fluorescence Imaging in a Single Microscope: Biomedical Applications Using a Synchrotron Infrared Source (Invited)"*; Rev. Sci. Instrum. **2002**; 73 (3): 1357–1360. DOI:10.1063/1.1435824.

Mitri E., A. Pozzato, G. Coceano, D. Cojoc, L. Vaccari, M. Tormen, G. Grenci *"Highly IR-Transparent Microfluidic Chip with Surface-Modified $BaF_2$ Optical Windows for Infrared Microspectroscopy of Living Cells."*; Microelectron. Eng. **2013**;107:6–9. DOI: 10.1016/j.mee.2013.02.068

Moss, D.A., M. Keese, R. Pepperkok *"IR Microspectroscopy of Live Cells"*; Vib. Spectrosc. **2005**, 38, 185–191 DOI: 10.1016/j.vibspec.2005.04.004

Moss, D.A. (Ed.) *"Biomedical Applications of Synchrotron Infrared Microscopy"* RSC Publishing, **2010**, DOI:10.1039/9781849731997. Hardback ISBN: 0854041541 PDF ISBN: 1849731997

Nasse, M. J., S. Ratti, M. Giordano, C. J. Hirschmugl "*Demountable Liquid/Flow Cell for in Vivo Infrared Microspectroscopy of Biological Specimens*"; *Appl. Spectrosc.* **2009**, 63, 10, 1181. DOI: 10.1366/000370209789553101

Naumann, D. "*FT-Infrared and Raman Spectroscopy in Biomedical Research*"; *Appl. Spectrosc. Rev.* **2001**, 36:2-3, 239-298, DOI: 10.1081/ASR-100106157

Noda, I. *"Generalized Two-Dimensional Correlation Spectroscopy"* in *Frontiers of Molecular Spectroscopy* J. Laane (Ed.), Elsevier **2009**, Chapter 13 367-381, ISBN 9780444531759, DOI: 10.1016/B978-0-444-53175-9.00013-1.

Noda, I., Y. Ozaki *"Two-Dimensional Correlation Spectroscopy"* **2004** John Wiley &Sons, Chichester, England. Print ISBN:9780471623915 |Online ISBN:9780470012406 |DOI:10.1002/0470012404

Nickell, S., P. S.-H. Park, W. Baumeister, K. Palczewski "*Three-Dimensional Architecture of Murine Rod Outer Segments Determined by Cryoelectron Tomography*"; *J. Cell. Biol.* **2007**; 177 (5): 917–925. DOI: 10.1083/jcb.200612010

Quaroni, L., T. Zlateva, D. Bedolla, S. Massaro, V. Torre *"Measurement of Molecular Orientation in a Subcellular Compartment by Synchrotron Infrared Spectromicroscopy."*; *ChemPhysChem* **2008**, 9: 1380-1382. DOI: 10.1002/cphc.200800211

Quaroni, L., E. Normand *"Two-Dimensional Correlation Spectroscopy Analysis for the Recovery of Weak Bands from Time-Resolved Infrared Spectra of Single Cells."* AIP Conf. Proc. **2010**; 1214 (1): 66–68. DOI: 10.1063/1.3326352

Quaroni, L., T. Zlateva *"Infrared Spectromicroscopy of Biochemistry in Functional Single Cells"* Analyst **2011a**, 136, 3219-3232 DOI: 10.1039/c1an15060j

Quaroni, L., T. Zlateva, E. Normand "*Detection of Weak Absorption Changes from Molecular Events in Time-Resolved FT-IR Spectromicroscopy Measurements of Single Functional Cells*"; *Anal. Chem.* **2011b**, 83, 19, 7371–7380. DOI: 10.1021/ac201318z

Quaroni, L., F. Zobi *"Cellular Imaging with Metal Carbonyl Complexes"* in *Inorganic Chemical Biology* G. Gasser (Ed.) **2014a** DOI:10.1002/9781118682975.ch5

Quaroni, L., T. Zlateva "*Real-Time Metabolic Analysis of Living Cancer Cells with Correlated Cellular Spectro-microscopy*"; *Anal. Chem.* **2014b**, 86, 14, 6887–6895. DOI: 10.1021/ac501561x

Quaroni L., T. Zlateva, B. Sarafimov, H. W. Kreuzer, K. Wehbe, E. L. Hegg, G. Cinque *"Synchrotron Based Infrared Imaging and Spectroscopy Via Focal Plane Array on Live Fibroblasts in $D_2O$ Enriched Medium"*; *Biophys. Chem.* **2014c,** 189, 40-48. DOI: 10.1016/j.bpc.2014.03.002

Quaroni, L., M. Obst, M. Nowak, F. Zobi "*Three-Dimensional Mid-Infrared Tomographic Imaging of Endogenous and Exogenous Molecules in a Single Intact Cell with Subcellular*

arXiv:2511.04143

Resolution"; *Angew. Chem., Int. Ed. Engl.* **2015**, 54, 1, 318-322. DOI: 10.1002/anie.201407728

Quaroni, L., T. Zlateva, K. Wehbe, G. Cinque *"Infrared Imaging of Small Molecules in Living Cells: From in Vitro Metabolic Analysis to Cytopathology"; Faraday Discuss.* **2016**, 187, 259-271. DOI: 10.1039/c5fd00156k

Quaroni, L. *"Infrared Microscopy in the Study of Cellular Biochemistry"; Infrared Phys. Technol.* **2020a,** 105, 102779. DOI: 10.1016/j.infrared.2018.11.026

Quaroni, L., I. Benmessaoud, B. Vileno, E. Horváth, L. Forró *"Infrared and 2-Dimensional Correlation Spectroscopy Study of the Effect of $CH_3NH_3PbI_3$ and $CH_3NH_3SnI_3$ Photovoltaic Perovskites on Eukaryotic Cells."; Molecules* **2020b**, 25, 336. DOI: 10.3390/molecules25020336

Santoro, G., T. Zlateva, A. Ruggi, L. Quaroni, F. Zobi *"Synthesis, Characterisation and Cellular Location of Cytotoxic Constitutional Organometallic Isomers of Rhenium Delivered on a Cyanocobalamin Scaffold"; Dalton Trans.* **2015**, 44, 6999-7007. DOI: 10.1039/c4dt03598d.

Salzer, R., H. W. Siesler (Eds.) *"Infrared and Raman Spectroscopic Imaging"* **2009**, Wiley-VCH Verlag GmbH & Co. KGaA Print ISBN:9783527319930 |Online ISBN:9783527628230 |DOI:10.1002/9783527628230

Shurvell, H.F. *"Spectra-Structure Correlations in the Mid- and Far- Infrared"* in *Handbook of Vibrational Spectroscopy* J. Chalmers, P. Griffiths (Eds.). John Wiley & Son, Chichester, **2002** 3, 1783-1816. ISBN-10 0471988472; ISBN-13 978-0471988472.

Siebert, F. *"Infrared Spectroscopy Applied to Biochemical and Biological Problems"; Methods Enzymol.* **1995**, 246, 501-526, Academic Press, London. ISBN: 0076-6879

Sommer A.J. *"Mid-Infrared Transmission Microspectroscopy"* in *Handbook of Vibrational Spectroscopy* J. Chalmers, P. Griffiths (Eds.) John Wiley & Son, Chichester, **2002** 2, 1369-1385. ISBN-10 0471988472; ISBN-13 978-0471988472.

Tamm L.K., S.A. Tatulian *"Infrared Spectroscopy of Proteins and Peptides in Lipid Bilayers."; Q. Rev. Biophys.* **1997**; 30(4): 365-429. DOI: 10.1017/S0033583597003375

Tobin, M.J; L. Puskar, R.L. Barber, E.C. Harvey, P. Heraud, B.R. Wood, K.R. Bambery, C.T. Dillon, K.L. Munro *"FTIR Spectroscopy of Single Live Cells in Aqueous Media by Synchrotron IR Microscopy Using Microfabricated Sample Holders."; Vib. Spectrosc.* **2010**, 53, 1, 34-38, DOI: 10.1016/j.vibspec.2010.02.005.

Vaccari, L., G. Birarda, L. Businaro, S. Pacor, G. Grenci *"Infrared Microspectroscopy of Live Cells in Microfluidic Devices (MD-IRMS): Toward a Powerful Label-Free Cell-Based Assay"; Anal. Chem.* **2012** 84 (11), 4768-4775 DOI: 10.1021/ac300313x

Vannocci, T., R. Notario Manzano, O. Beccalli, B. Bettegazzi, F. Grohovaz, G. Cinque, A. De Riso, L. Quaroni, F. Codazzi, A. Pastore *"Adding a Temporal Dimension to the Study of Friedreich's Ataxia: The Effect of Frataxin Overexpression in a Human Cell Model."; Dis. Models Mech.* **2018**; 11 (6): dmm032706. DOI: 10.1242/dmm.032706.

Vannocci, T., L. Quaroni, A. de Riso, G. Milordini, M. Wolna, G. Cinque, A. Pastore "*Label-Free, Real-Time Measurement of Metabolism of Adherent and Suspended Single Cells by In-Cell Fourier Transform Infrared Microspectroscopy*"; Int. J. Mol. Sci. **2021**, 22, 19: 10742. DOI: 10.3390/ijms221910742

Venyaminov, S. Yu., F. G. Prendergast "*Water ($H_2O$ and $D_2O$) Molar Absorptivity in the 1000–4000 $cm^{-1}$ Range and Quantitative Infrared Spectroscopy of Aqueous Solutions*"; Anal. Biochem. **1997**, 248, 234–245. DOI: 10.1006/abio.1997.2136

Wehbe, K., J. Filik, M.D. Frogley, G. Cinque "*The Effect of Optical Substrates on Micro-FTIR Analysis of Single Mammalian Cells.*"; Anal. Bioanal. Chem. **2013**, 405, 1311–1324. DOI: 10.1007/s00216-012-6521-6

Wetzel, D.L., S.M. Levine "*Imaging Molecular Chemistry with Infrared Microscopy*"; Science **1999**, 285, 5431, 1224-1225. DOI: 10.1126/science.285.5431.1224.

Wilson, E.B., J.C. Decius, P.C. Cross "*Molecular Vibrations: The Theory of Infrared and Raman Spectra*" **1980** Dover Publications ISBN: 048663941X

Zhao, R., L. Quaroni, A. G. Casson "*Fourier Transform Infrared (FTIR) Spectromicroscopic Characterization of Stem-Like Cell Populations in Human Esophageal Normal and Adenocarcinoma Cell Lines.*" Analyst **2010**, 135, 1: 53-61. DOI: 10.1039/B914311D

Zobi, F., L. Quaroni, G. Santoro, T. Zlateva, O. Blacque, B. Sarafimov, M.C. Schaub, A. Yu. Bogdanova "*Live-Fibroblast IR Imaging of a Cytoprotective PhotoCORM Activated with Visible Light.*"; J. Med. Chem. **2013**, 56, 17: 6719-6731. DOI: 10.1021/jm400527k